\documentclass[]{aa}

\usepackage[varg]{txfonts}
\usepackage{graphicx}
\usepackage{multirow}
\usepackage{ulem}
\usepackage[verbose=true]{hyperref}
\bibpunct{(}{)}{;}{a}{}{,}

\begin{document}

\title{Gas absorption and dust extinction towards the Orion Nebula Cluster}

\author{Birgit Hasenberger\inst{\ref{inst1}} 
\and Jan Forbrich\inst{\ref{inst1}, \ref{inst2}} 
\and Jo\~{a}o Alves \inst{\ref{inst1}}
\and Scott J. Wolk \inst{\ref{inst2}}
\and Stefan Meingast \inst{\ref{inst1}}
\and Konstantin V. Getman \inst{\ref{inst3}}
\and Ignazio Pillitteri \inst{\ref{inst2}, \ref{inst4}}}

\institute{Department for Astrophysics, University of Vienna, Türkenschanzstra\ss e 17, 1180 Vienna, Austria \label{inst1} 
\and Harvard-Smithsonian Center for Astrophysics, 60 Garden Street, Cambridge, MA 02138, USA \label{inst2}
\and Department of Astronomy and Astrophysics, 525 Davey Laboratory, Pennsylvania State University, University Park, PA 16802, USA \label{inst3}
\and INAF-Osservatorio Astronomico di Palermo, Piazza del Parlamento 1, 90134, Palermo, Italy \label{inst4}}

\authorrunning{B. Hasenberger et al.}
\titlerunning{Gas absorption and dust extinction towards the ONC}

\date{Received <date> /
Accepted <date>}

\abstract{}{We characterise the relation between the gas and dust content of the interstellar medium towards young stellar objects in the Orion Nebula Cluster.}
{X-ray observations provide estimates of the absorbing equivalent hydrogen column density $N_H$ based on spectral fits. Near-infrared extinction values are calculated from intrinsic and observed colour magnitudes $(J-H)$ and $(H-K_s)$ as given by the VISTA Orion A survey. A linear fit of the correlation between column density and extinction values $A_V$ yields an estimate of the $N_H/A_V$ ratio. We investigate systematic uncertainties of the results by describing and (if possible) quantifying the influence of circumstellar material and the adopted extinction law, X-ray models, and elemental abundances on the $N_H/A_V$ ratio.}
{Assuming a Galactic extinction law with $R_V=3.1$ and solar abundances by \citet{Anders1989}, we deduce an $N_H/A_V$ ratio of $(1.39 \pm 0.14) \cdot 10^{21}\ \mathrm{cm}^{-2}\ \mathrm{mag}^{-1}$ for Class III sources in the Orion Nebula Cluster where the given error does not include systematic uncertainties. This ratio is consistent with similar studies in other star-forming regions and approximately 31\% lower than the Galactic value. We find no obvious trends in the spatial distribution of $N_H/A_V$ ratios. Changes in the assumed extinction law and elemental abundances are demonstrated to have a relevant impact on deduced $A_V$ and $N_H$ values, respectively. Large systematic uncertainties associated with metal abundances in the Orion Nebula Cluster represent the primary limitation for the deduction of a definitive $N_H/A_V$ ratio and the physical interpretation of these results.}{}


\maketitle

\section{Introduction}
\label{sec:Intro}

The characteristics and evolution of the interstellar medium (ISM) are closely connected to major astrophysical questions in areas such as molecular cloud evolution, star and planet formation, and galactic evolution. Undoubtedly, knowledge of the properties of ISM constituents, i.e. gas and dust, and in particular the relation between them is vital when characterising interstellar environments. In this study, we describe the relationship between gas column density $N_H$ and dust extinction $A_V$ towards young stellar objects (YSOs) in the \object{Orion Nebula Cluster} (ONC). Simplistically, measurements of these quantities towards the same object are expected to be correlated since matter is traced for the same distance along the same line of sight and the gas and dust components are assumed to coincide spatially. The ratio $N_H/A_V$ is frequently used synonymously with the gas-to-dust mass ratio, but although the quantities are related, the conversion is not straightforward. A number of factors influence this relation, such as dust grain properties and elemental abundances, which in general are difficult to constrain observationally and thus necessitate several additional assumptions. The assumptions made during data analysis and the interpretation of observational results need to undergo careful consideration in order to ensure the reliability of deduced physical properties, e.g. $N_H/A_V$ and the gas-to-dust mass ratio.

In the Milky Way, previous studies using multiple independent methods have found a uniform $N_H/A_V$ ratio ${\sim}2 \cdot 10^{21}$~cm$^{-2}$~mag$^{-1}$ \citep[e.g. ][]{Bohlin1978, Predehl1995, Watson2011} and a gas-to-dust mass ratio ${\sim}100$ for regions dominated by diffuse interstellar matter throughout the sky. Towards young, star-forming regions, observed $N_H/A_V$ ratios are typically lower than or similar to the value in the diffuse ISM \citep[e.g. ][]{Vuong2003, Winston2010, Pillitteri2013}. The origin of this difference is unclear, although the literature provides several possible explanations: altered grain properties in cold, dense environments caused by, for example, grain growth \citep[e.g. ][]{Goldsmith1997, RomanDuval2014}, lower metal abundances in star-forming regions \citep{Vuong2003}, and a lower gas-to-dust mass ratio \citep[e.g. ][]{Bohlin1978, Vuong2003}.

The ONC provides excellent conditions for this study despite observational difficulties arising from nebulosity and high, non-uniform extinction in the region as well as its complex three-dimensional structure \citep[see e.g. ][]{ODell2001}. At a distance of 414\,pc \citep{Menten2007} it is the nearest massive star-forming region, making it a prime target for numerous observational studies. The wealth and high quality of available data in various wavelength ranges are crucial ingredients for data analysis since they allow for a statistical evaluation of results, the constraining of stellar parameters, and the selection of highly reliable data points from a large sample.

\section{Data}
\label{sec:data}

\begin{figure}
  \resizebox{\hsize}{!}{\includegraphics{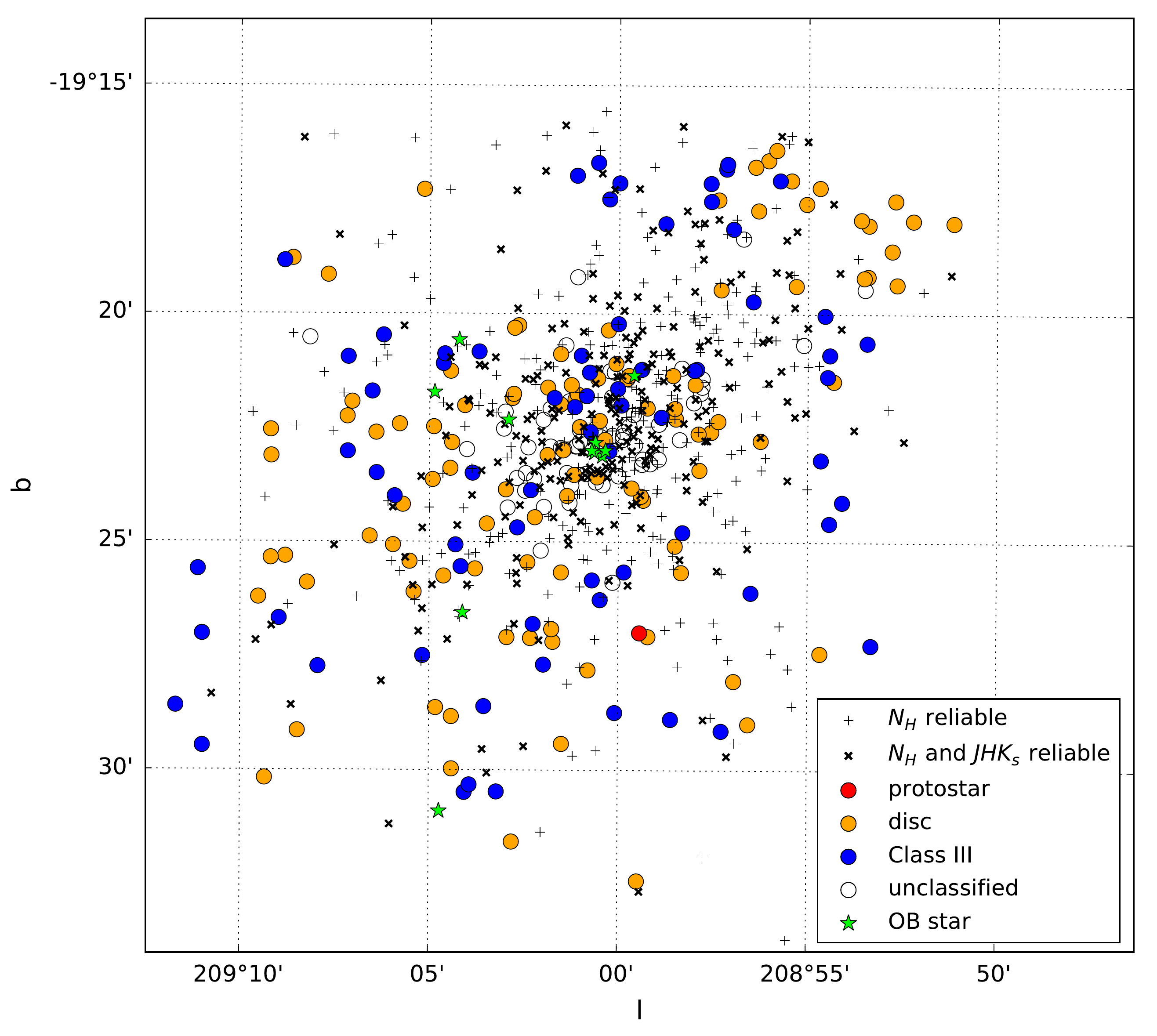}}
  \caption{Spatial distribution of objects in the sample in galactic coordinates. Green stars indicate the positions of the brightest OB stars in the region. Filled circles correspond to sources with an identified spectral class and reliable $N_H$ and $A_V$ values.}
  \label{fig:sky}
\end{figure}

This study utilises observations in the X-ray range from the \textit{Chandra} Orion Ultradeep Project (COUP), in the near-infrared (NIR) from the VISTA Orion A survey, and in the mid-infrared (MIR) from the \textit{Spitzer} Orion survey. While the gas column density is frequently derived from L$\alpha$ absorption at UV wavelengths, X-ray observations provide an alternative method for inferring this quantity. In the X-ray range, photoelectric absorption dominates the observed extinction. Consequently, the absorption cross section is characterised by cut-offs corresponding to the energies of electronic transitions and is determined by the column density of elements in the absorbing medium. If a set of elemental abundances is assumed, the equivalent hydrogen column density can thus be deduced from observations of X-ray spectra. This approach provides certain advantages over UV observations, as summarised by \citet{Vuong2003}: X-ray absorption is capable of tracing the total hydrogen column density since it is not sensitive to the physical or chemical state of absorbing elements and probes material to higher extinctions. These characteristics are particularly relevant for the investigation of the ONC region presented here. In contrast, extinction due to dust is deduced from NIR observations by comparison with the stars' intrinsic colours, requiring the assumption of an extinction law. Data in the MIR regime are used for YSO classification. Since circumstellar material can possibly alter the derived $N_H/A_V$ ratio (see Sect. \ref{subsubsec:circ}), the classification of YSOs and in particular a distinction between IR-excess and discless sources is essential for this study. In the following sections, all the employed data sets are described in more detail along with further data processing steps. The spatial distribution of objects is shown in Fig.~\ref{fig:sky}.

\subsection{COUP}
\label{subsec:COUP}

The COUP catalogue \citep{Getman2005} is based on \textit{Chandra} ACIS observations from 2003 with a total exposure time of approximately ten days. It provides X-ray properties of 1616 point sources including spectral fitting results. \citet{Getman2005} used a one- or two-component thermal plasma emission model (MEKAL code as implemented in Xspec, based on models by, among others, \citealt{Mewe1985} and \citealt{Liedahl1995}) modified by a single absorption component (Wabs code as implemented in Xspec, based on cross-sections by \citet{Morrison1983}) to model the X-ray spectra. Solar abundances were adopted from \citet{Anders1989} and stellar coronal metal abundances were assumed to be proportional to the Sun's photospheric abundances by a factor of 0.3 for all sources (see discussion of this value in Sect.~\ref{subsubsec:abund}). The relevant fitting parameter for this study is the equivalent hydrogen column density $N_H$. The influence of the adopted X-ray model components and solar abundance values on the derived column density is discussed in Sect. \ref{subsubsec:model} and \ref{subsubsec:abund}. As an estimate of the error in $N_H$, the statistical 1-$\sigma$ error from the fit and a constant value of $10^{21}$~cm$^{-2}$ are added. This constant factor is introduced so that uncertainties, in particular for low column densities, are more accurately reflected, as these typically show unrealistically small statistical errors considering the difficulty of determining $N_H$ values based on the low-energy end of X-ray spectra. In order to only include sources with reliable $N_H$ values, several groups of objects are excluded from the sample: Sources which are flagged in the COUP catalogue as having exhibited problems during the fit (non-empty entry in the \textit{sFlags} column, which is Column~17 in Table~6 of \citet{Getman2005}) or as providing only an upper limit on the column density, and sources with a reduced $\chi^2$ value of less than 0.5 or larger than 2.0 are excluded. The resulting sample contains 728 sources.

\subsection{VISTA Orion A survey}
\label{subsec:VISION}

As part of VISION (Vienna Survey in Orion), the VISTA Orion~A survey covers $\approx 18.3$\,deg$^2$ of the sky spanning the entire Orion~A molecular cloud \citep{Meingast2016}. The authors created a new reduction pipeline in order to optimise spatial resolution, and produced images as well as a source catalogue in the $JHK_s$ NIR bands containing almost 800\,000 objects. Complications due to the highly variable background during source extraction near the Orion Nebula are addressed and described in the survey paper by \citet{Meingast2016}.

We chose the matching radius between the COUP and VISION catalogues based on the investigation of a histogram of separations between X-ray and NIR sources up to $3\arcsec$ taking only the closest match for each COUP source into account. With a bin size of $0.1\arcsec$ we find a peak in the number of matched sources between $0.2\arcsec$ and $0.3\arcsec$. The matching radius is chosen as the separation at which the number of matched sources is equal to 1\% of the peak value. We thus employ a matching radius of $1\arcsec$, yielding a sample of 1292 COUP objects with NIR counterparts, 655 of which with reliable $N_H$ values. The sample is then restricted to sources with magnitude measurements in all three bands, ensuring that extinction values are deduced from two valid NIR colours for all objects. Additionally, we compare the VISION data to the $JHK_s$ values given in the COUP catalogue, which include data obtained by the NIR camera ISAAC at the ESO Very Large Telescope for the inner quarter of the COUP field \citep{Getman2005}. The comparison generally shows excellent agreement between the two data sets. Any sources with magnitude differences larger than 1\,mag in any band are excluded from the sample. Thus, 465 objects with reliable data in both the X-ray and NIR regime constitute the correlated sample.

Dust extinction values are deduced from NIR colour magnitudes using the NICER algorithm \citep{Lombardi2001}, which estimates $A_V$ based on intrinsic and observed NIR colours and an assumed extinction law. Here, we adopt the extinction law as given by \citet{Cardelli1989} with a ratio of total to selective extinction $A_V/E(B-V)$, denoted as $R_V$, of 3.1. Implications of higher $R_V$ values are discussed in Sect. \ref{subsubsec:ext}. A set of intrinsic colours, $(J-H)_0$ and $(H-K)_0$, is deduced for each individual source by employing information on their spectral type: Source positions are correlated with an optical survey of the ONC by \citet{Hillenbrand2013}, providing the spectral type for each matched source. Intrinsic NIR colours are then obtained by associating spectral types with empirically derived intrinsic colours as given by \citet{Scandariato2012}. For 239 sources with reliable X-ray and NIR data, intrinsic colours can be deduced with this method. These objects constitute our final sample. An uncertainty of 2 subclasses in the spectral type and an additional error of 0.03 accounting for the uncertainty of the colour magnitudes for each spectral type are assumed. The final products are dust extinction values $A_V$ and their corresponding uncertainties, which are calculated by taking both photometric errors and uncertainties of intrinsic colours into account.

\subsection{\textit{Spitzer} Orion survey}
\label{subsec:Spitzer}

\begin{figure}
  \resizebox{\hsize}{!}{\includegraphics{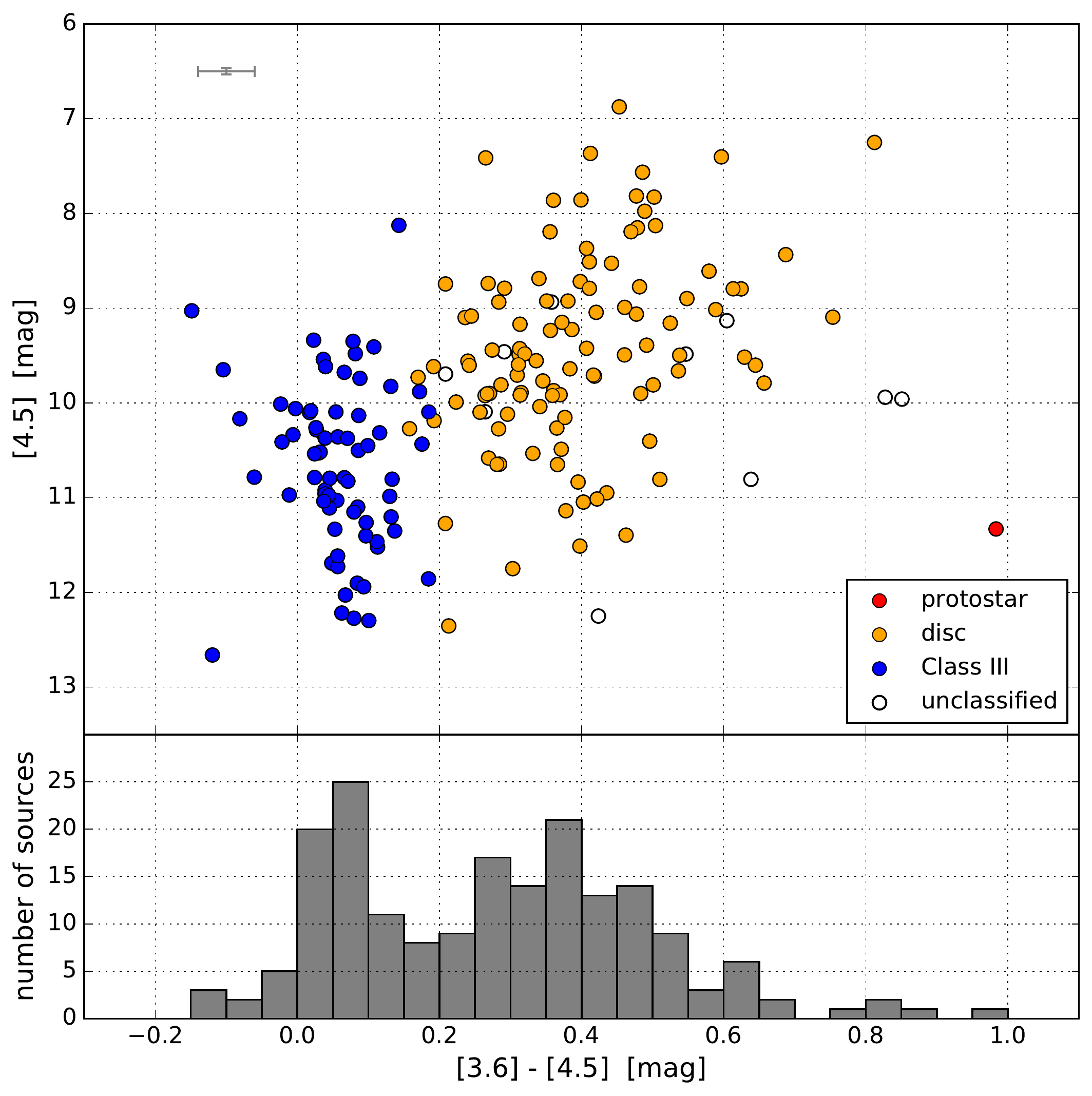}}
  \caption{Diagrams used to identify Class III objects in the final sample. \textit{Upper panel:} Colour-magnitudes. For 90\% of sources the uncertainties are lower than those represented by the error bars in the upper left corner of the figure. Median errors are smaller than the symbol size. \textit{Lower panel: } Colour magnitudes $[3.6]-[4.5]$.}
  \label{fig:class}
\end{figure}

A survey of the Orion~A and B clouds was conducted by the \textit{Spitzer Space Telescope} \citep{Megeath2012}, mapping the regions in five bands within the MIR regime. The point source catalogue produced from these observations contains almost 300\,000 sources which have been classified according to procedures based on those described by \citet{Gutermuth2009} and \citet{Kryukova2012}. This source classification scheme is aimed at identifying a reliable sample of IR-excess sources by applying cuts in colour-colour and colour-magnitude diagrams. As it does not attempt to identify Class III sources, an additional cut is necessary. In this study, the objects in the final sample show a bimodal distribution in the difference between the observed magnitudes in the 3.6~$\mu$m and 4.5~$\mu$m IRAC wavebands, $[3.6]-[4.5]$ (see Fig.~\ref{fig:class}). We thus use this quantity to separate IR-excess from discless sources: All objects which are not flagged as an IR-excess source by \citet{Megeath2012} and exhibit a colour magnitude $[3.6]-[4.5] < 0.2$ are considered discless sources. Since all objects in the final sample show X-ray emission, we identify all discless sources within this sample as Class III YSOs. In the final sample of 239 objects, we find a counterpart in the \textit{Spitzer} point source catalogue for 195 sources, 106 of which are flagged as disc stars and one as a protostar. Following the procedure described above, we identify 70 sources as Class III objects, leaving a total of 62 YSOs unclassified. Of these, 44 are not classified because no match with a \textit{Spitzer} source is found, 8 owing to missing measurements in the $[3.6]$ and/or $[4.5]$ waveband, and 10 YSOs were not classified as IR-excess sources by \citet{Megeath2012} while not fulfilling the criteria of a Class III YSO. Throughout the analysis, the entire sample (including unclassified sources) and two subsamples including exclusively either IR-excess or Class III sources are used. A catalogue of objects in the final sample along with the main properties used in this work is made available at the CDS. Table~\ref{tab:cat} lists the column names and descriptions of this catalogue.

\begin{table*}
\caption{Description of columns in the catalogue of the final sample available at the CDS.}
\label{tab:cat}
\centering
\begin{tabular}{c l l}
\hline\hline
Column number & Column name & Description \\
\hline
1  & COUP\_no & COUP identification number \\
2  & RA & RA from the COUP catalogue \\
3  & dec & dec from the COUP catalogue \\
4  & NH & $N_H$ from the COUP catalogue [cm$^{-2}$] \\
5  & NH\_err & 1-$\sigma$ error of $N_H$ as assumed in this work [cm$^{-2}$] \\
6  & VISION\_no & VISION identification number \\
7  & J & $J$ magnitude from VISION [mag] \\
8  & J\_err & 1-$\sigma$ error of $J$ magnitude from VISION [mag] \\
9  & H & $H$ magnitude from VISION [mag] \\
10 & H\_err & 1-$\sigma$ error of $H$ magnitude from VISION [mag] \\
11 & Ks & $K_s$ magnitude from VISION [mag] \\
12 & Ks\_err & 1-$\sigma$ error of $K_s$ magnitude from VISION [mag] \\
13 & sptype & Spectral type as deduced from \citet{Hillenbrand2013} \\
14 & JH0 & Intrinsic $(J-H)$ magnitude as deduced from \citet{Scandariato2012} [mag] \\
15 & JH0\_err & 1-$\sigma$ error of intrinsic $(J-H)$ magnitude [mag] \\
16 & HK0 & Intrinsic $(H-K_s)$ magnitude as deduced from \citet{Scandariato2012} [mag] \\
17 & HK0\_err & 1-$\sigma$ error of intrinsic $(H-K_s)$ magnitude [mag] \\
18 & AV & $A_V$ for $R_V=3.1$\tablefootmark{a} as deduced in this work [mag] \\
19 & AV\_err & 1-$\sigma$ error of $A_V$ for $R_V=3.1$ as deduced in this work [mag] \\
20 & class & Classification (p -- protostar, d -- disc star, cl3 -- Class III source, uncl -- unclassified) \\
\hline
\end{tabular}
\tablefoot{\tablefoottext{a}{$A_V$ for $R_V=5.5$ can be calculated by multiplication with the factor 1.18 (see Sect.~\ref{subsubsec:ext}).}}
\end{table*}

\section{Results}
\label{sec:res}

\begin{figure}
  \resizebox{\hsize}{!}{\includegraphics{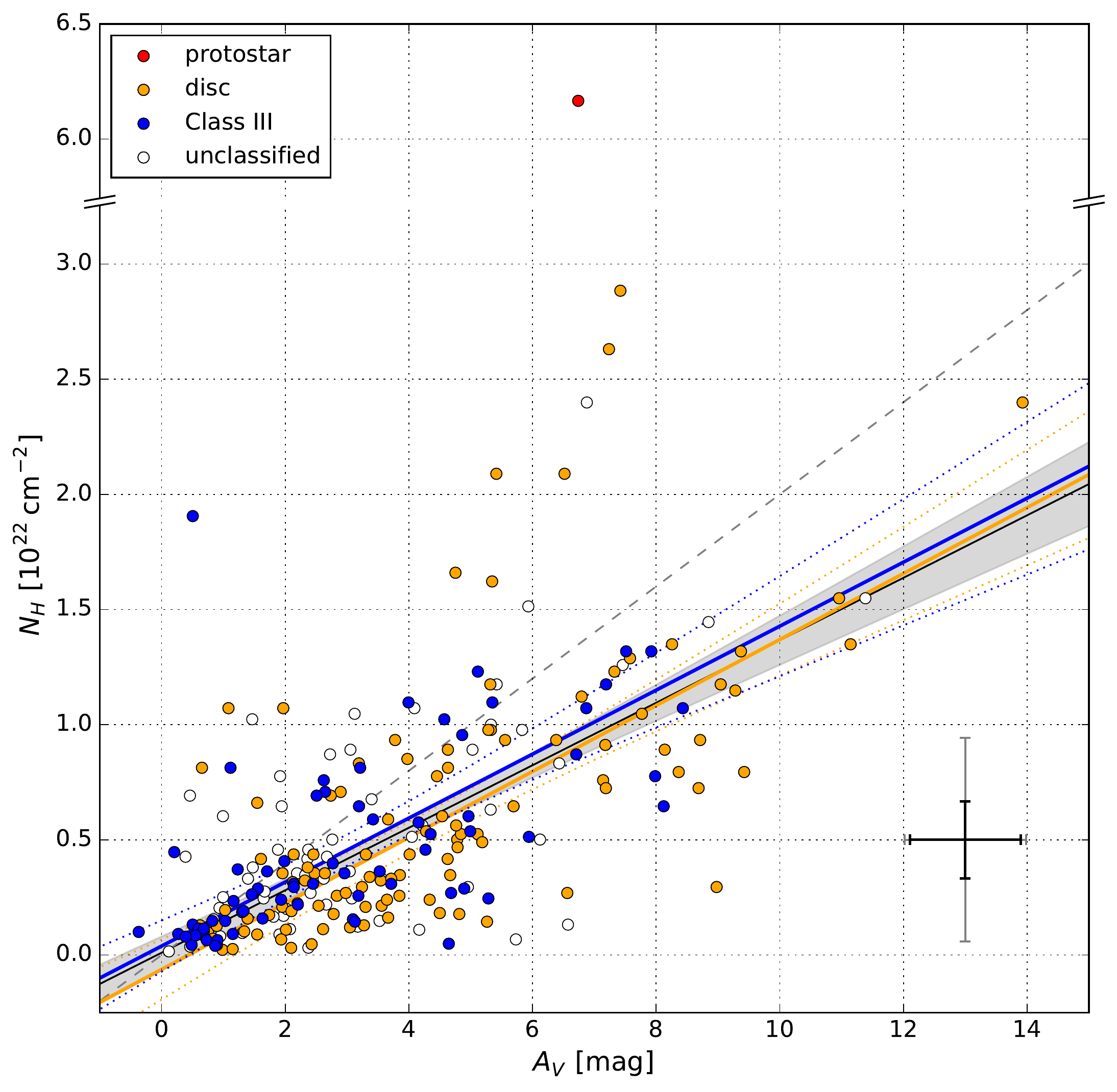}}
  \caption{Equivalent hydrogen column density $N_H$ versus extinction in the $V$-band $A_V$. For 90\% of sources the uncertainties are lower than those represented by the grey error bars in the lower right corner of the figure. Median values of errors are indicated by the black error bars. The solid black, orange, and blue lines are linear fits to the entire sample, IR-excess sources only, and Class III objects only, respectively. The shaded area and dotted orange or blue lines indicate the 1-$\sigma$ confidence band of the fits. The dashed grey line corresponds to the $N_H/A_V$ ratio of the diffuse ISM in the Milky Way, $N_H = 2 \cdot 10^{21}$~cm$^{-2}$~mag$^{-1} \cdot A_V$.}
  \label{fig:NH_AV}
\end{figure}

A comparison of hydrogen column density and extinction for the final sample shows, upon a first visual inspection, a distinct correlation between the two parameters with some scatter (see Fig.~\ref{fig:NH_AV}). The trend suggests an $N_H/A_V$ ratio below the value for the diffuse ISM in the Milky Way with the majority of data points ($76\%$) below the $N_H/A_V = 2 \cdot 10^{21}$~cm$^{-2}$~mag$^{-1}$ relation. Considering different YSO evolutionary classes, no significant difference in the slope of the relation or magnitude of scatter can be found between IR-excess and Class III sources. However, a noticeable group of data points well above the Galactic relation can be observed between extinction values of ${\sim}5$ and 8. Except for two unclassified YSOs, all other objects within this group are identified as disc sources. Furthermore, the single protostar within the sample has an exceptionally high column density of ${\sim}6 \cdot 10^{22}$~cm$^{-2}$~mag$^{-1}$ at $A_V\sim7$. Possible explanations for this behaviour of protostars and YSOs with discs are discussed in Sect.~\ref{subsubsec:circ}. In addition to astrophysical reasons, inferring high column densities from X-ray fits can be caused by the degeneracy of the fit with respect to the plasma temperature and $N_H$. The best-fit model might thus underestimate the plasma temperature and overestimate the column density. In particular, this phenomenon is relevant for objects with flares which may harden the X-ray spectrum and may be associated with higher plasma temperatures. We inspected the COUP light curves \citep{Getman2005} of the outlying sources with considerably higher $N_H/A_V$ ratios and found that several showed possible signs of flaring during the observations. Nevertheless, not all YSOs in this group exhibit flares, necessitating other explanations for their high column density values. In the case of the object classified as a protostar, the VISION images reveal a second source at a small projected distance of ${\sim}1.4''$ that is not part of the COUP catalogue. The unusual $N_H/A_V$ ratio of the protostar in the final sample might thus be a result of confusion with or contamination by this second source.

We now quantify the relation between column densities and extinction values for three samples, namely the entire sample, IR-excess sources only, and Class III sources only, by calculating linear fits to the data points. An orthogonal distance regression routine (ODRPACK as implemented in the Python library Scipy, \citet{Jones2001}) is employed to enable the consideration of errors in both column density and extinction. The fits yield the following relations
\begin{equation*}
\begin{aligned}
N_H^{(\mathrm{all})} =\ &(1.36 \pm 0.08) \cdot 10^{21}\ \mathrm{cm}^{-2}\ \mathrm{mag}^{-1} \cdot A_V \\
&+ (0.10 \pm 0.34) \cdot 10^{21}\ \mathrm{cm}^{-2},
\end{aligned}
\end{equation*}
\begin{equation*}
\begin{aligned}
N_H^{(\mathrm{IRex})} =\ &(1.43 \pm 0.14) \cdot 10^{21}\ \mathrm{cm}^{-2}\ \mathrm{mag}^{-1} \cdot A_V \\
&+ (-0.63 \pm 0.68) \cdot 10^{21}\ \mathrm{cm}^{-2},\ \mathrm{and}
\end{aligned}
\end{equation*}
\begin{equation*}
\begin{aligned}
N_H^{(\mathrm{III})} =\ &(1.39 \pm 0.14) \cdot 10^{21}\ \mathrm{cm}^{-2}\ \mathrm{mag}^{-1} \cdot A_V \\
&+ (0.39 \pm 0.47) \cdot 10^{21}\ \mathrm{cm}^{-2},
\end{aligned}
\end{equation*}
for the entire sample, IR-excess sources only, and Class III sources only, respectively. Given uncertainties correspond to statistical 1-$\sigma$ errors from the fit. The intercept is included as a fit parameter in order to account for any systematic offsets that could occur, e.g. due to incorrect estimation of intrinsic colours. Since the best-fit values for the intercept are compatible with zero within their 1-$\sigma$ uncertainties for all three samples, the data are unlikely to suffer from such offsets.  The fits confirm a lower ratio of $N_H/A_V$ compared to the value in the diffuse ISM by approximately 31\%, and that different YSO classes exhibit similar best-fit parameters and uncertainties. The numerical results reported here are deduced from column densities taken directly from the COUP catalogue, which assumes solar abundance values by \citet{Anders1989}. The $N_H/A_V$ ratio based on revised abundances by \citet{Asplund2009} can be found in Sect. \ref{subsubsec:abund}.

\begin{figure}
  \resizebox{\hsize}{!}{\includegraphics{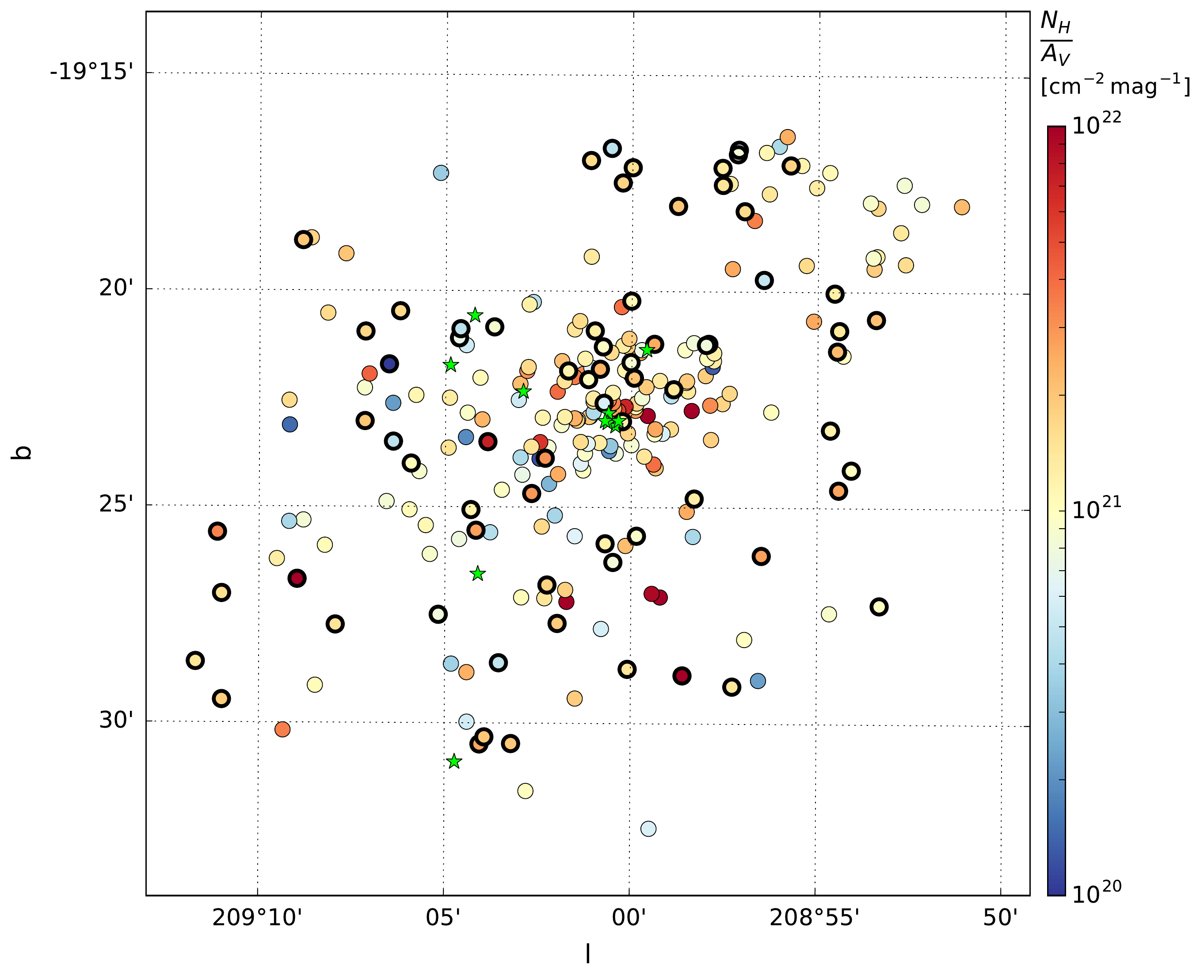}}
  \caption{Spatial distribution of sources in the final sample in galactic coordinates, colour-coded by their individual $N_H/A_V$ value. Class III sources are indicated by a thicker symbol outline. Green stars indicate the positions of the brightest OB stars in the region. No obvious trends in the distribution of $N_H/A_V$ values can be observed.}
  \label{fig:NHAV_sky}
\end{figure}

The spatial distribution of $N_H/A_V$ values calculated for individual sources is shown in Fig.~\ref{fig:NHAV_sky}. No obvious global gradients or clusters of stars with similar ratios can be found.

\begin{figure}
  \resizebox{\hsize}{!}{\includegraphics{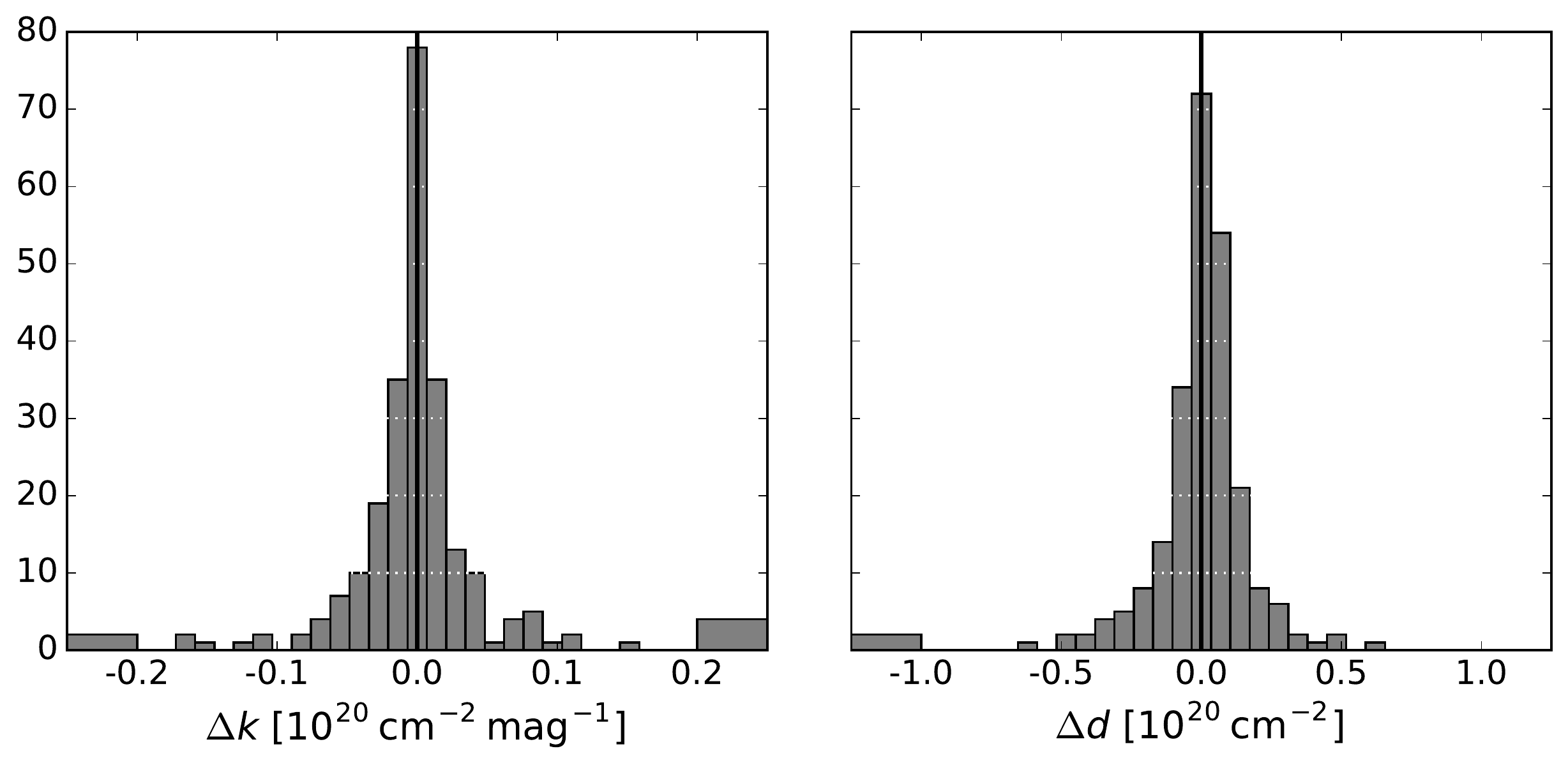}}
  \caption{Difference between best-fit parameters for the entire sample and the sample reduced by one source. The first and last bins contain all data points below the lowest or above the largest labelled value on the x-axis, respectively. \textit{Left panel:} Slope of the linear function $k$. \textit{Right panel:} Intercept of the linear function $d$.}
  \label{fig:boot_indi}
\end{figure}

\begin{figure}
  \resizebox{\hsize}{!}{\includegraphics{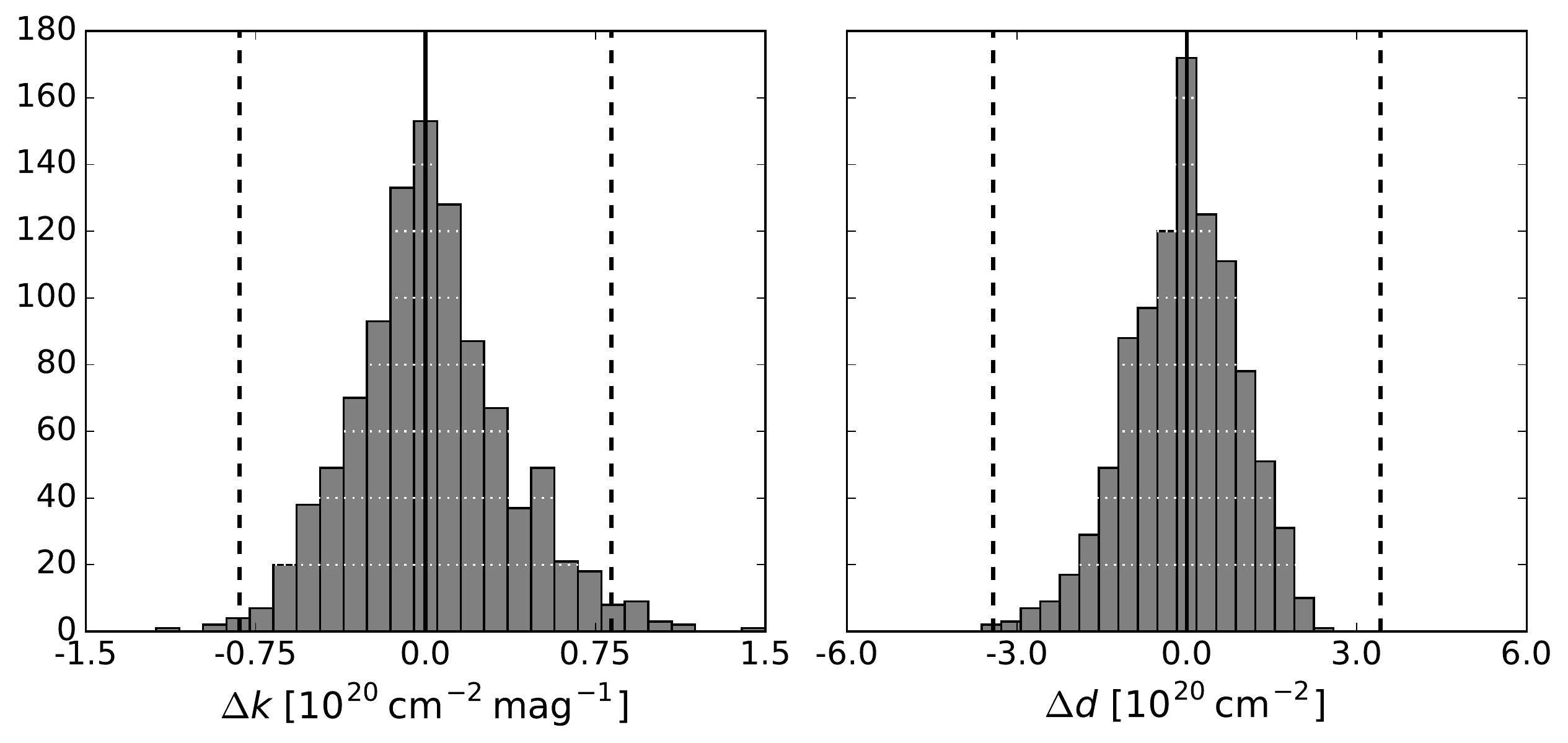}}
  \caption{Same as Fig.~\ref{fig:boot_indi}, but for the sample reduced by 10\% of the total number of objects in the final sample. There are no simulations for which the refitted parameters assume values beyond the x-axis limits. Dashed lines indicate statistical 1-$\sigma$ uncertainties of the fit to the entire sample.}
  \label{fig:boot_10p}
\end{figure}

To test the robustness of the fits, two experiments are conducted. In the first experiment, we employ the jackknife method by removing a single source from the sample and refitting the reduced data set using the same routine as described earlier. This is done for every source in the sample and the resulting slope and intercept values are measured. In the second experiment, a random choice of sources constituting 10\% of the full sample are removed before refitting the remaining data points. The results of 1000 simulations of this kind are obtained. Removing individual sources from the sample, all objects cause changes of the order of a few $10^{19}$~cm$^{-2}$~mag$^{-1}$ or less in the derived slope (see Fig.~\ref{fig:boot_indi}). For 95\% of objects, the difference in the slope from the fit to the entire sample is less than $1.1 \cdot 10^{19}$~cm$^{-2}$~mag$^{-1}$. Even in the case of removing 10\% of sources from the data set, the distribution of parameters drops quickly as one moves away from the best-fit value for the full sample (see Fig.~\ref{fig:boot_10p}): In 95\% of all simulations, the resulting slope deviates by less than $6.8 \cdot 10^{19}$~cm$^{-2}$~mag$^{-1}$. We thus conclude that the fits are not dominated by a small subset of the sample, but correctly represent the general trend contained within the data.

\section{Discussion}
\label{sec:disc}

\subsection{Comparison to previous studies}
\label{subsec:prev}

\begin{table*}
\caption{Selection of previous studies employing X-ray observations to derive $N_H/A_V$ in star-forming regions or the diffuse ISM of the Milky Way}
\label{tab:NHAV}
\centering
\begin{tabular}{l c c l l}
\hline\hline
\multirow{2}{*}{Target region} & $N_H/A_V$ & Number of & \multirow{2}{*}{Comments} & \multirow{2}{*}{Reference} \\
 & [cm$^{-2}$ mag$^{-1}$] & data points & &\\
\hline
\multirow{5}{*}{diffuse ISM} & $2.2 \cdot 10^{21}$ & $<638$\tablefootmark{a} & X-ray absorption towards supernova remnants& \citet{Watson2011} \\
 & \multirow{2}{*}{$1.8 \cdot 10^{21}$} & \multirow{2}{*}{29} & X-ray scattering halos around point sources & \multirow{2}{*}{\citet{Predehl1995}} \\
 & & & and supernova remnants &  \\
 & $2.2 \cdot 10^{21}$ & 7 & X-ray absorption towards diffuse sources & \citet{Gorenstein1975} \\
 & $2.2 \cdot 10^{21}$ & 5 & X-ray absorption towards supernova remnants & \citet{Ryter1975}\tablefootmark{b} \\
\hline
\object{L1641} & $0.8 \cdot 10^{21}$ & 211 & includes 133 Class III YSOs & \citet{Pillitteri2013}\tablefootmark{c} \\
\object{IRAS 20050+2720} & $1.2 \cdot 10^{21}$ & 21 & includes 10 Class III YSOs & \citet{Guenther2012} \\
\object{NGC 1333} & $1.0 \cdot 10^{21}$ & 26 & includes 7 Class III YSOs & \citet{Winston2010} \\
Serpens & $0.7 \cdot 10^{21}$ & $<60$\tablefootmark{d} & includes $<21$\tablefootmark{d} Class III YSOs & \citet{Winston2007} \\
\object{$\rho$ Oph} & $1.5 \cdot 10^{21}$ & 22 & sample consists entirely of Class III YSOs& \citet{Vuong2003} \\
\object{RCW 38} & $1.7 \cdot 10^{21}$ & 43 & includes 16 Class III YSOs& \citet{Winston2011} \\
\object{RCW 108} & $2.0 \cdot 10^{21}$ & 117 & Class III YSOs not discussed separately & \citet{Wolk2008} \\
ONC & $1.4 \cdot 10^{21}$ & 239 & includes 70 Class III YSOs & this work \\
\hline                                   
\end{tabular}
\tablefoot{\tablefoottext{a}{The authors do not report the number of data points used in their analysis, but state that they included 638 Gamma-ray bursts in their sample, while the data quality for "a significant fraction" of objects is not sufficient to derive column densities.}
\tablefoottext{b}{The listed $N_H/A_V$ ratio has been calculated from the $N_H/E(B-V)$ value reported in the paper by assuming $R_V=3.1$.}
\tablefoottext{c}{The listed $N_H/A_V$ ratio has been calculated from the $N_H/A_K$ value reported in the paper by assuming $A_K/A_V = 0.11$ as given by \citet{Cardelli1989} for $R_V=3.1$.}
\tablefoottext{d}{The authors do not report the number of data points used in their analysis, therefore the stated number includes sources with insufficient X-ray counts for the derivation of column densities.}}
\end{table*}

Table~\ref{tab:NHAV} summarises key information on previous studies of the $N_H/A_V$ ratio in the diffuse ISM and various star-forming regions. The comparability of these studies is limited, however, owing to the large differences in sample size and consideration of systematic effects. Regarding the sample size, this work is among the largest studies of X-ray absorption and dust extinction in a star-forming region. The importance of large sample sizes for these comparisons is also noted by \citet{Vuong2003}, who refrain from reporting $N_H/A_V$ ratios for several nearby star-forming regions, including Orion, owing to the small number of data points in their final sample.

Despite deviations in data analysis procedures, it is evident from the reported $N_H/A_V$ values that the ISM typically exhibits a ratio larger than or equal to those found in star-forming regions. This effect is commonly explained by grain growth in the cold, dense, and dark environment of molecular clouds which leads to higher extinction values \citep[e.g. ][]{Carrasco1973, Cardelli1988}. \citet{Vuong2003} argue, however, that it is possible to attribute the lower $N_H/A_V$ ratio towards $\rho$~Oph purely to a difference in metallicity with respect to the ISM. In addition, there appears to be a trend for low-mass star-forming regions such as Serpens \citep{Winston2007}, NGC~1333 \citep{Winston2010}, or L1641 \citep{Pillitteri2013} to show lower $N_H/A_V$ ratios than regions which harbour one or more OB stars such as RCW~108 \citep{Wolk2008} or RCW~38 \citep{Winston2011}. This phenomenon has been associated with ablation of dust grains by UV radiation from massive stars \citep{Pillitteri2013}.

For the ONC, the $N_H/A_V$ ratio derived here is lower than in the diffuse ISM and is consistent with values typically found in star-forming regions. Following the hypothesis of grain growth in molecular clouds and destruction of large grains by UV radiation as mentioned above, sources further away from OB stars should show lower $N_H/A_V$ values. The main contributors to the UV flux in the ONC are the Trapezium stars; however, no obvious radial gradient in $N_H/A_V$ originating from these stars can be found. Still, this experiment is not capable of falsifying the hypothesis as several critical aspects of the influence of UV radiation on dust grains are not considered (in part because they are unknown), including the three-dimensional arrangement of stars within the cluster and the attenuation of UV radiation within the medium it pervades.

\subsection{Investigation of assumptions and complications}
\label{subsec:assum}

During the data analysis in this study, a number of assumptions have to be made in order to estimate the amount of gas and dust along the line of sight, namely the extinction law towards the source, X-ray emission and absorption models, and an associated set of elemental abundances. While commonly neglected in studies of gas absorption and dust extinction, the influence of these assumptions on the final result is evaluated in this work by considering a different shape of the extinction law, more recent X-ray spectral models, and an update of solar abundance values.

As both hydrogen column density and dust extinction probe material all along the line of sight up to the location of the source, variations in the properties of the pervading matter along the line of sight additionally affect the measurement. In the case that gas and dust are not well-mixed, as suggested by \citet{Hopkins2016} for certain properties of the medium, $N_H$ and $A_V$ would not correlate for some sources.  Similarly, X-ray and NIR observations towards the same source could trace slightly different lines of sight as emission in these two regimes has distinct physical and thus possibly also spatial origins. The correlation between $N_H$ and $A_V$ might be lost if the lines of sight are not identical and therefore different media are probed. In addition, the COUP and VISION observations were not conducted simultaneously, introducing uncertainties due to variability into the analysis. Since YSOs in earlier evolutionary classes are more variable than Class-III sources in both the NIR and X-ray regime \citep{Flaccomio2012, Rice2015}, variability might be more problematic for the analysis of IR-excess sources in our sample. All of the mentioned effects could be the origin of some of the scatter in the comparison of $N_H$ and $A_V$, and are generally difficult to avoid or quantify. Nevertheless, observations in the MIR allow for the evolutionary classification of objects and thus enable the removal of sources which are likely to possess significant amounts of circumstellar material such as envelopes or discs.

\subsubsection{Circumstellar material}
\label{subsubsec:circ}

In the literature, a number of different mechanisms associated with circumstellar material are considered as possible reasons for a change in the $N_H/A_V$ ratio. Several are connected to a change in grain properties in the vicinity of the star, for example grain growth and planet formation. Furthermore, \citet{Dullemond2003} investigated the influence of scattering off dust grains in circumstellar discs on the emerging spectrum and demonstrate that this effect is relevant in the NIR and MIR regimes. Taking both gravitational settling of grains in the disc and grain growth into account, \citet{Rettig2006} and \citet{Horne2012} show that the observed gas-to-dust ratio is significantly lower for discs which are seen edge-on. For a given source, it is difficult to predict which of the mentioned effects are relevant and to what extent they influence the observed $N_H/A_V$ ratio. However, Class III sources, possessing little or no circumstellar material, are not affected by these processes and are thus more reliable tracers of the interstellar $N_H/A_V$ ratio.

In an attempt to avoid the contribution of circumstellar material to derived extinction values, a different method could be applied. Instead of dereddening NIR colours of YSOs to their intrinsic colours using the NICER algorithm, it is possible to deredden sources along a given extinction vector to a locus of intrinsic colours in the NIR colour-colour diagram. For sources with discs, this locus would correspond to empirically derived colours, e.g. by \citet{Meyer1997} for classical T Tauri stars (CTTS) in Taurus, which would ideally produce $A_V$ values unaffected by the circumstellar disc. However, there are a number of uncertainties introduced by this approach: First, assuming a CTTS locus for the ONC would require the locus to be derived from a region of stars with similar properties, such as spectral types, stellar radii, or disc accretion rates \citep{Meyer1997}. Investigating whether a CTTS locus from the literature is appropriate for stars in the ONC and the systematic errors arising from this effect is challenging because of the various stellar and disc parameters which influence the locus. Second, the CTTS locus is associated with uncertainties due to the scatter of intrinsic colours around this line in the colour-colour diagram, adding to the overall systematic error in derived extinction values. Third, the comparability of NIR and X-ray observations might be compromised if $A_V$ values are corrected for the effects of circumstellar material, while $N_H$ values are not. As a result, we do not deredden disc sources to a CTTS locus and employ one method to derive $A_V$ values consistently for all YSO classes. The unknown effect of circumstellar material on extinction and hydrogen column density measurements is accounted for by excluding IR-excess sources from the sample used to derive the $N_H/A_V$ ratio. We therefore expect the value derived solely from data for Class III objects to be the most representative measurement of this ratio in the interstellar medium towards YSOs in the ONC.

Previous studies which also used YSO classification to avoid objects possibly surrounded by discs or envelopes report diverse relations between the overall $N_H/A_V$ ratio of IR-excess and Class III sources. Several authors find higher values for IR-excess sources than Class III YSOs \citep{Winston2007, Winston2010, Wolk2010}, while the opposite is the case, for example in the L1641 region \citep{Pillitteri2013}. \citet{Guenther2012} do not detect a difference in $N_H/A_V$ between IR-excess and Class III sources. For the ONC, the $N_H/A_V$ values deduced by fitting the sample of IR-excess and Class III objects separately are consistent within their errorbars. The largest occurring deviation in $N_H$ within the extinction range studied here ($A_V=0$ to 14) is only ${\sim}10^{21}$~cm$^{-2}$. As noted by \citet{Guenther2012}, possible explanations include the following: The extinction law could be similar for interstellar and circumstellar matter or any material surrounding the source might not contribute significantly to the measured extinction value since $A_V$ is dominated by cloud material rather than circumstellar matter. 

Although no substantial difference in $N_H/A_V$ is found between the IR-excess and Class III source samples, the evolutionary class can still provide crucial information on the position of individual objects in the $N_H$-versus-$A_V$ diagram. In particular, a conspicuous group of outliers with high values of $N_H/A_V$ consists almost entirely of disc sources, suggesting that disc material has a significant impact on the $N_H/A_V$ ratio for these objects. Values of $N_H/A_V$ that are considerably higher than predicted by the Galactic relation have also been found for a sample of nearby classical T-Tauri stars \citep{Guenther2008}. Findings from the literature and this study therefore highlight the importance of including YSO classification in studies of the ISM when analysing lines of sight towards YSOs. Discs and envelopes around YSOs represent environments in which grain growth can occur \citep{Kwon2009, Ricci2010, Testi2014}. Models by \citet{Ormel2011} suggest that as grains grow to $\mu$m sizes the NIR colour excess increases, resulting in higher extinction values, while it decreases if grains become larger than ${\sim}10\,\mu$m. Furthermore, the inclination of the circumstellar disc may alter the observed ratio between gas absorption and dust extinction: \citet{Horne2012} find that this value can be higher than the interstellar ratio by factors of up to several tens for discs which are not seen edge-on. The literature thus provides at least two possible astrophysical causes for the exceptionally high $N_H/A_V$ ratios we deduce for a number of IR-excess sources in our sample, namely dust grains of very large ($\gtrsim 10\,\mu$m) sizes, and an effect of the viewing angle on the disc.

\subsubsection{Extinction law}
\label{subsubsec:ext}

\begin{table}
\caption{Selection of studies deriving $R_V$ towards the ONC}
\label{tab:RV}
\centering
\begin{tabular}{c l}
\hline\hline
$R_V$ & Reference \\
\hline
3.0 & \citet{Walker1969} \\
\multirow{2}{*}{3.1} & \citet{Penston1975}\tablefootmark{a}, \\
 & \citet{DaRio2010} \\
3.7 - 5.4\tablefootmark{b} & \citet{Johnson1967} \\
4.8 & \citet{Mendez1967} \\
5.23 - 5.5\tablefootmark{b} & \citet{Cardelli1989} \\
\multirow{3}{*}{5.5} & \citet{Costero1970}, \\
 & \citet{Mathis1981}, \\
 & \citet{DaRio2015} \\
$\geq 5.5$ & \citet{Lee1968} \\
\hline
\end{tabular}
\tablefoot{\tablefoottext{a}{The authors report a normal extinction law for the ONC but do not explicitly state a value for $R_V$.}\tablefoottext{b}{The range represents the minimum and maximum $R_V$ value the authors found for the observed stars.}}
\end{table}

The shape of the extinction law can be characterised by the parameter $R_V$, which is dependent on dust grain properties such as the size distribution, shape, or composition \citep[e.g. ][]{Mathis1989, Kim1994, Weingartner2001, Fitzpatrick2007}. While these dependencies are generally complex, it has been established that larger grains lead to higher values of $R_V$. For the diffuse ISM in the Milky Way, $R_V=3.1$. Up to this point, all calculations have been conducted assuming the same $R_V$ for the ONC. However, there is no clear consensus on the value of $R_V$ in the ONC; several authors support a value close to the $R_V$ of the diffuse ISM while others find an anomalous $R_V$ of ${\sim}5.5$ (see Table \ref{tab:RV}). An extinction law different from the Galactic relation towards stars in the ONC was observed almost 80 years ago by \citet{Baade1937}, who suggested that this phenomenon is related to an altered grain size distribution. In regions of cold and dense material such as molecular clouds, $R_V$ values different from the Galactic value are now commonly attributed to grain growth in these environments \citep[e.g. ][]{Carrasco1973, Cardelli1988, Draine1990, Foster2013}. In order to assess the influence of the adopted extinction law on our results, the analysis is repeated assuming $R_V=5.5$. Owing to the steeper slope of the extinction law in the NIR range, $A_V$ values are decreased leading to an $N_H/A_V$ ratio higher by ${\sim}18\%$. For Class III sources only,
\begin{equation*}
\begin{aligned}
N_H^{(\mathrm{III})} =\ &(1.62 \pm 0.16) \cdot 10^{21}\ \mathrm{cm}^{-2}\ \mathrm{mag}^{-1} \cdot A_V \\
&+ (0.42 \pm 0.47) \cdot 10^{21}\ \mathrm{cm}^{-2}.
\end{aligned}
\end{equation*}

The $N_H/A_V$ ratio towards the ONC is then closer to, but still clearly below the Galactic relation. An anomalous extinction law found previously for the ONC is thus not sufficient to entirely explain the difference in $N_H/A_V$ between the diffuse ISM in the Milky Way and the ONC region. Nevertheless, this experiment shows that changes in grain properties are able to account for a part of the discrepancy, and that therefore a lower-than-Galactic $N_H/A_V$ ratio does not necessarily imply a lower gas-to-dust mass ratio. Considering the intricate connections between dust grain characteristics and $R_V$, and the uncertainty associated with the value of $R_V$, no firm conclusions on grain properties are possible in this context.

\subsubsection{X-ray models}
\label{subsubsec:model}

Spectral fit data reported in the COUP catalogue are based on X-ray fits of thermal plasma emission (MEKAL) and absorption (Wabs) models which have been revised recently. The APEC emission model \citep{Smith2001} incorporates an extended database of atomic transitions and the Phabs absorption model uses updated extinction cross-sections by \citet{Balucinska-Church1992} and \citet{Yan1998}. In addition, the adopted solar abundances cannot be altered for the Wabs model, which invariably assumes the set of abundances by \citet{Anders1989}. This implies that the influence of abundances on the deduced column density values (see Sect. \ref{subsubsec:abund}) cannot be investigated using the Wabs absorption model. For both the Wabs and Phabs model, the elemental abundances of the absorbing medium are fixed to the given solar values. The X-ray emitting plasma, however, can be set to have abundances proportional to the Sun's by a factor given as a parameter in the MEKAL and APEC models.

In order to asses the changes in $N_H$ introduced by using revised X-ray models, the grouped COUP spectra are refitted with one or two APEC components modified by a single Phabs absorption term. Initial values for the absorbing column density as well as the temperature of the plasma emission are set to the corresponding best-fit values given in the COUP catalogue. The number of emission components and assumed abundance values are also adopted from the catalogue.

\begin{figure}
  \resizebox{\hsize}{!}{\includegraphics{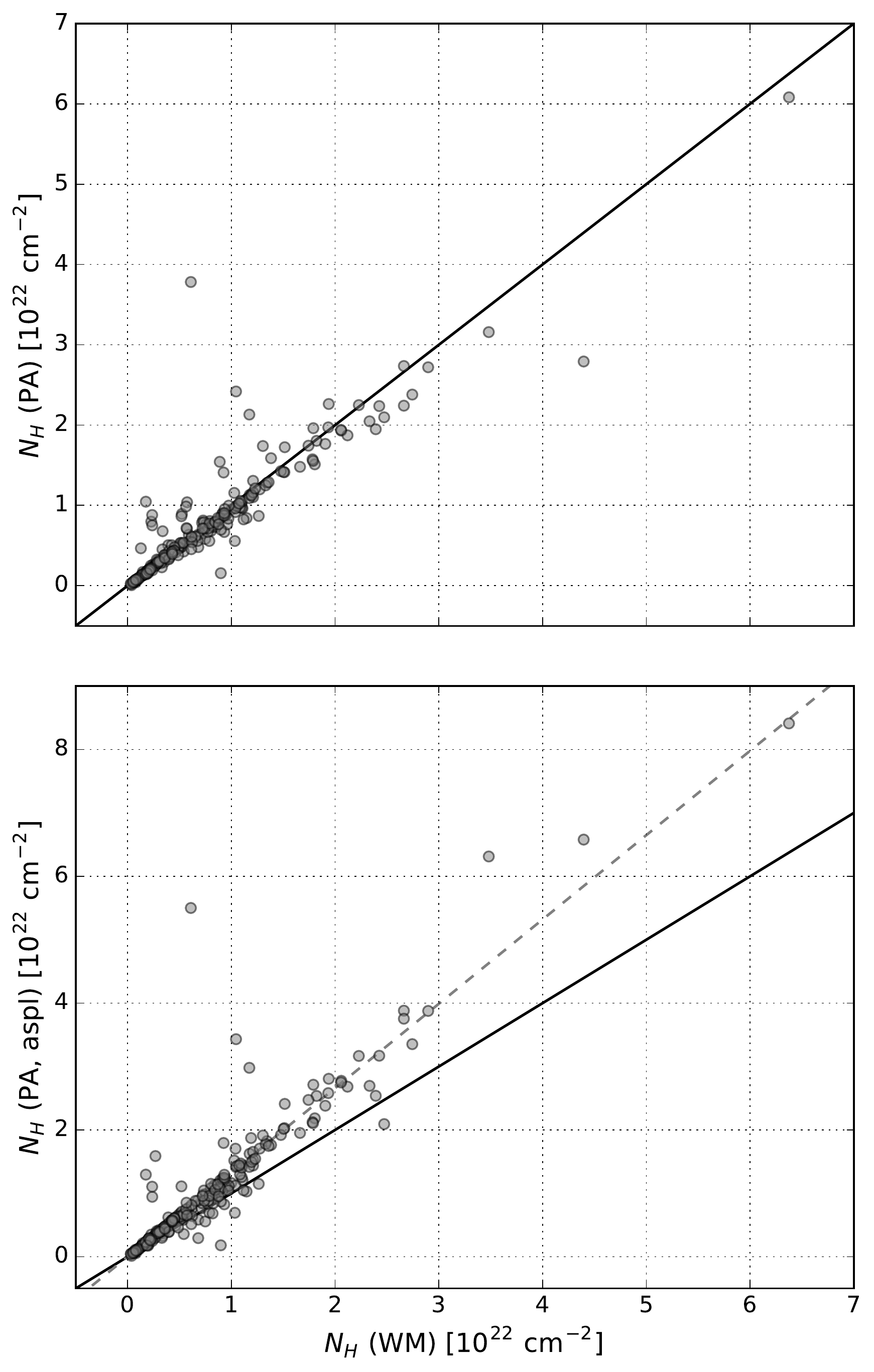}}
  \caption{Equivalent hydrogen column densities calculated using Phabs and APEC (PA) versus Wabs and MEKAL (WM) models. Solid lines indicate equality. \textit{Upper panel:} For both PA and WM, abundances by \citet{Anders1989} are adopted. \textit{Lower panel:} For PA, abundances by \citet{Asplund2009} are adopted, and for WM by \citet{Anders1989}. The dashed grey line is intended as a guide to the eye and corresponds to $1.33 \cdot N_H$ (WM).}
  \label{fig:PA_WM}
\end{figure}

Generally, hydrogen column densities derived with Wabs-MEKAL and Phabs-APEC models are compatible with each other (see Fig. \ref{fig:PA_WM}, upper panel). The revised models systematically produce slightly lower $N_H$ values, an effect that can be ascribed to the difference between the absorption models. It is typically small compared to statistical errors or systematic uncertainties stemming from, e.g. the adopted extinction law or elemental abundances. The majority of sources for which the derived $N_H$ values do not agree well are characterised by high values of $\chi^2$ when using the Phabs and APEC models, indicating that the fit may not represent the global minimum of $\chi^2$. These outliers are therefore not considered as evidence of systematic trends. We thus conclude that the revision of X-ray models does not lead to a substantial change in derived $N_H$ values in the context of this study.

A recent study by \citet{Corrales2016}, however, shows that fitting X-ray spectra with absorption models such as Phabs leads to overestimated column density values as these models do not account for X-ray scattering. The authors model and fit \textit{Chandra} spectra assuming that radiation is scattered by a homogeneous ISM between the observer and the source, and other properties typical for the observation of Galactic X-ray binaries. From this analysis, they find that $N_H$ values are overestimated by at least 25\% if models without scattering are used. In order to estimate the relevance of X-ray scattering for YSOs in the ONC, a detailed model, for example of dust properties and the spatial arrangement of the X-ray source and the scattering medium, is required. Since this is beyond the scope of this study, no statement on a possible systematic error introduced by the neglect of X-ray scattering can be made. In all cases, this effect cannot be the cause of the lower $N_H/A_V$ ratio towards the ONC with respect to the diffuse ISM in our Galaxy since including scattering in our X-ray analysis could only reduce derived $N_H$ values.

\subsubsection{Elemental abundances}
\label{subsubsec:abund}

The assumption of elemental abundances is required to calculate line flux levels and interpret X-ray absorption, which is caused primarily by heavier elements such as C, N, and O, as an equivalent hydrogen column density. Abundances in the Orion Nebula have been shown to be similar to the Sun's \citep{Esteban2004}, justifying the use of solar values for this region. Coronally active stars, however, exhibit lower metal abundances than the Sun owing to the inverse FIP effect (see e.g. reviews by \citet{Guedel2009, Laming2015}). This phenomenon can be accounted for by assuming a metallicity < 1 for stellar coronal abundances in the spectral emission model. During the fitting process of the COUP data, \citet{Getman2005} adopted solar abundance values as given by \citet{Anders1989} and a metallicity of 0.3 for the X-ray emitting plasma. In the following, we discuss the implications of these two assumptions.

Since the study by \citet{Anders1989}, elemental abundances have been revised for both the ISM \citep{Wilms2000} and the Sun \citep{Asplund2009}. The effects of these modifications on deduced column densities are difficult to estimate a priori, although examples from the literature can be found. In a study of the gas-to-dust ratio towards $\rho$~Oph, \citet{Vuong2003} obtain $N_H$ values ${\sim}20\%$ higher when adopting updated ISM abundances by \citet{Wilms2000}. An increase of ${\sim}30\%$ in $N_H$ is reported if the recently revised solar abundances by \citet{Asplund2009} are assumed for X-ray studies of supernova remnants \citep{Foight2015}. We employ a similar approach as in the previous section and refit the COUP spectra using updated X-ray models, Phabs and APEC, and solar abundances by \citet{Asplund2009}. The stellar coronal metal abundance is fixed at 0.3 relative to solar for this experiment.

A comparison of $N_H$ values derived with different sets of abundances shows that the updated abundances produce column densities higher by ${\sim}33\%$ (see Fig. \ref{fig:PA_WM}, lower panel), consistent with the analysis by \citet{Foight2015}. We refit the relation between $N_H$ and $A_V$ in our final sample using these increased values by applying the same quality criteria on the X-ray fitting results (a reduced $\chi^2$ value higher than 0.5 and lower than 2.0) as for the fits in the COUP catalogue, and the same orthogonal distance regression algorithm as described in Sect.~\ref{sec:res}. For Class III sources only, the fit yields a slope of $(1.72 \pm 0.25) \cdot 10^{21}\ \mathrm{cm}^{-2}\ \mathrm{mag}^{-1}$ if an extinction law with $R_V = 3.1$ is assumed, and $(2.04 \pm 0.30) \cdot 10^{21}\ \mathrm{cm}^{-2}\ \mathrm{mag}^{-1}$ if $R_V = 5.5$. In both cases, the uncertainties are larger than if abundances by \citet{Anders1989} are assumed. This might be a consequence of the smaller sample size, namely 158 objects with 42 classified as Class III sources, due to the additional fit quality criterion we imposed. Although the $N_H/A_V$ ratios derived when assuming the revised solar abundances are systematically higher, the 1-$\sigma$ confidence bands overlap. In addition, they are compatible with the Galactic $N_H/A_V$ ratio, and appear to challenge our previous findings. However, studies of the gas-to-dust ratio in the diffuse ISM using X-rays also require the assumption of elemental abundances, and the corresponding value of $2.0 \cdot 10^{21}\ \mathrm{cm}^{-2}\ \mathrm{mag}^{-1}$ for the $N_H/A_V$ ratio was established before the revised solar abundance values. As demonstrated by \citet{Foight2015}, the assumption of updated abundances in analyses of the diffuse ISM would likewise produce a higher gas-to-dust ratio. The difference between the $N_H/A_V$ ratios towards the ONC and the diffuse ISM would thus remain unchanged if the assumptions on the proportionality of abundances to solar values hold.

For the emission component of X-ray model spectra, subsolar metal abundances are regularly assumed for YSOs in star-forming regions \citep{Imanishi2001, Feigelson2002, Getman2005, Guedel2007} since lower metallicities have been found for young stars, e.g. in $\rho$ Oph \citep{Kamata1997, Imanishi2001}. We convinced ourselves of the validity of this factor by comparing column densities derived when adopting a metallicity of 0.3 and 1.0. Solar abundances by \citet{Asplund2009} and updated X-ray models, Phabs and APEC, were used in both cases. The results show that such a change in metallicity does not appear to result in a systematic change in $N_H$, as both positive and negative differences occur at a given column density. However, a clear trend can be observed when comparing reduced $\chi^2$ values of the fits. Assuming subsolar metal abundances yields a lower reduced $\chi^2$ value for 86\% of sources in the final sample, suggesting that models adopting a metallicity of 1.0 systematically produce worse fits to the data than subsolar metallicities, confirming that a factor of 0.3 relative to solar abundances is a reasonable choice for our sample.

\begin{figure}
  \resizebox{\hsize}{!}{\includegraphics{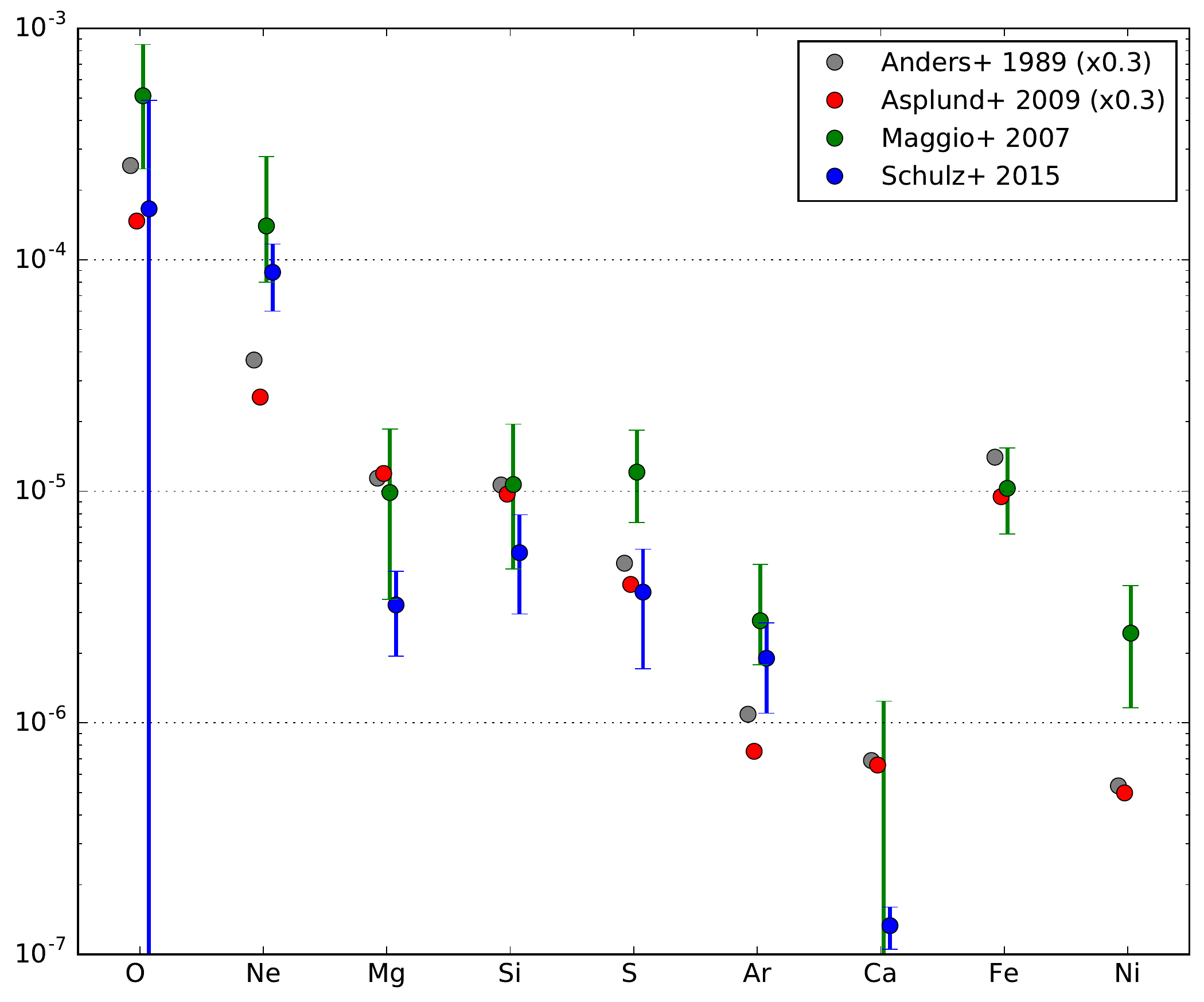}}
  \caption{Abundances as deduced by different authors for elements observed by \citet{Maggio2007}. Elements on the x-axis are arranged by atomic number. Error bars indicate 68\% and 90\% uncertainty limits for the results by \citet{Schulz2015} and \citet{Maggio2007}, respectively.}
  \label{fig:Abund}
\end{figure}

\citet{Vuong2003} present the possibility that the lower $N_H/A_V$ ratio in $\rho$~Oph compared to the Galactic value is caused entirely by lower metal abundances in this region. In order to investigate whether this scenario is also plausible for the ONC, the assumed abundances relative to solar values, i.e. as given by \citet{Anders1989} or \citet{Asplund2009}, are compared to recent results on abundances in the ONC. \citet{Esteban2004} find similar but slightly higher abundances in the Orion Nebula relative to solar values from the analysis of 555 emission lines at wavelengths between 3100 and 10\,400\,\AA. For the absorbing component of the X-ray model, the argument by \citet{Vuong2003} is thus not valid.

Considering the abundances in the X-ray emitting stellar atmospheres, \citet{Maggio2007} and \citet{Schulz2015} provide recent measurements for the ONC. As can be seen in Fig.~\ref{fig:Abund}, most elements exhibit large differences between the various studies. Abundances derived by \citet{Maggio2007} are in many cases consistent with or higher than the solar values, while the abundances by \citet{Schulz2015} lie mostly (although not exclusively) below these levels. We are thus unable to draw the overall conclusion that stellar coronal abundances corresponding to a metallicity of 0.3 relative to solar values overestimate the true metal abundances in YSOs in the ONC region since the observed abundances are not consistently lower than the values assumed in our analysis. Still, this argument cannot be considered sufficient to exclude the scenario suggested by \citet{Vuong2003} as this comparison of elemental abundances and any interpretation of it are complicated by the fact that the measured abundances are associated with relatively large error bars and that several elements critical for modelling X-ray spectra, such as C and N \citep{Wilms2000}, lack abundance measurements from observations of X-ray emitting stars in the ONC region altogether.

Overall, the elemental abundances can be identified as the most important contributor to systematic uncertainties considered in this study, since neither circumstellar material nor the choice of X-ray spectral models have a relevant effect on our results, and realistic changes in the assumed extinction law cause deviations of only ${\sim}18\%$ in $N_H/A_V$. Evidently, the precise measurement of metal abundances in the ONC is crucial for a reliable determination of the $N_H/A_V$ ratio in this region.

\section{Conclusions}
\label{sec:concl}

The relation between gas column density and dust extinction towards point sources in the ONC is investigated. Spectral fits to X-ray data are employed to deduce an equivalent hydrogen column density $N_H$, while NIR colour magnitudes and spectral classification allow for the calculation of dust extinction values $A_V$. A linear relation between the two parameters is found which suggests an $N_H/A_V$ ratio of $(1.39 \pm 0.14) \cdot 10^{21}\ \mathrm{cm}^{-2}\ \mathrm{mag}^{-1}$ for $R_V=3.1$, the given error representing statistical uncertainties only. This ratio is considerably lower than the ratio observed in the diffuse ISM and is consistent with values in other star-forming regions. Comparing the $N_H/A_V$ values deduced for samples of IR-excess and Class III objects only, no substantial difference is detected, indicating the limited influence of circumstellar material. The assumption of an anomalous extinction law with $R_V=5.5$ raises the $N_H/A_V$ ratio to $(1.64 \pm 0.16) \cdot 10^{21}\ \mathrm{cm}^{-2}\ \mathrm{mag}^{-1}$. Using updated X-ray models (Phabs and APEC instead of Wabs and MEKAL) does not produce significantly different column densities. Adopted elemental abundances, however, are shown to have a significant impact on deduced $N_H$ values while being poorly constrained towards the ONC. 

The size and quality of data sets used in this study have allowed us to derive an $N_H/A_V$ ratio with small statistical errors and to thoroughly investigate possible sources of systematic uncertainties. We find that systematic unknowns, in particular those associated with elemental abundances, introduce large uncertainties to our result which are difficult to control. Based on reasonable assumptions, an important statement can be made nonetheless: If the metal abundances in the ONC and the diffuse ISM are well described by solar values, the $N_H/A_V$ ratio in the ONC is lower than in the diffuse ISM by ${\sim}18 - 31\%$ (for $R_V=5.5-3.1$).

\begin{acknowledgements}
We thank the anonymous referee for comments and suggestions that helped improve the manuscript. Support for this work was provided by the National Aeronautics and Space Administration through Chandra Award Number GO2-13019X issued by the Chandra X-ray Observatory Center, which is operated by the Smithsonian Astrophysical Observatory for and on behalf of the National Aeronautics Space Administration under contract NAS8-03060. S.J.W. was supported by NASA contract NAS8-03060. K.V.G. acknowledges the support from the Chandra ACIS Team contract SV4-74018 (G. Garmire \& L. Townsley, PIs), issued by the Chandra X-ray Center. This research made use of Astropy, a community-developed core Python package for Astronomy \citep{Astropy2013}; matplotlib, a Python library for publication quality graphics \citep{Hunter2007}; SciPy \citep{Jones2001}; and XSPEC \citep{XSPEC1996}. 
\end{acknowledgements}

\bibliographystyle{aa}
\bibliography{mybib}

\end{document}